\documentclass[useAMS,usenatbib]{mn2e}

\usepackage{times}

\usepackage{multicol}
\usepackage{graphicx}
\usepackage{amssymb}
\usepackage{amsfonts}
\usepackage{amsbsy}
\usepackage{graphicx}
\def\MBH{M_{\rm BH}}
\def\MCh{M_{\rm Ch}}
\def\Rt{R_t}
\def\dM{\delta M}
  \def\tloss{\delta t}
\def\dR{\Delta}
\def\rmin{r_{\min}}
\def\tacc{t_{\rm acc}}
\def\simlt{\lower.5ex\hbox{$\; \buildrel < \over \sim \;$}}
\def\simgt{\lower.5ex\hbox{$\; \buildrel > \over \sim \;$}}

\title[White dwarfs stripped by massive black holes]
      {White dwarfs stripped by massive black holes: sources of coincident gravitational and electromagnetic radiation}
\author[I. Zalamea, K. Menou, A. M. Beloborodov]
{I. Zalamea$^{1}$, K. Menou$^{1,2}$ and 
A. M. Beloborodov$^{1,3}$ \\
$^{1}$Physics Department, Astronomy Department and Columbia Astrophysics Laboratory,
Columbia University, New York, NY 10027, USA\\
$^{2}$Kavli Institute for Theoretical Physics, UCSB, Santa Barbara, CA 93106-4030\\
$^{3}$Astro-Space Center of Lebedev Physical 
Institute, Profsojuznaja 84/32, Moscow 117810, Russia
}
\begin{document}
\date{Accepted ---. Received ---; in original form ---}
\pagerange{\pageref{firstpage}--\pageref{lastpage}} \pubyear{---}
\maketitle
\label{firstpage}
\begin{abstract}
White dwarfs inspiraling into black holes of mass $\MBH\simgt
10^5M_\odot$ are detectable sources of gravitational waves in the LISA
band.  In many of these events, the white dwarf begins to lose mass
during the main observational phase of the inspiral.  The mass loss
starts gently and can last for thousands of orbits.  The white dwarf
matter overflows the Roche lobe through the $L_1$ point at each
pericenter passage and the mass loss repeats periodically.  The
process occurs very close to the black hole and the released gas can
accrete, 
creating a bright source of radiation with
luminosity close to the Eddington limit, $L\sim 10^{43}$~erg~s$^{-1}$.
This class of inspirals offers a promising scenario for dual
detections of gravitational waves and electromagnetic radiation.
\end{abstract}
%
%
\section{Introduction}
One of the goals of the Laser Interferometer Space Antenna (LISA)
mission is to detect gravitational waves from compact stellar objects
spiraling into massive black holes, a class of events called
extreme-mass-ratio inspirals (EMRIs) (e.g., Hils \& Bender 1995; Gair
et al. 2004; Barack \& Cutler 2004; see Hughes 2009 and Sathyaprakash
\& Schutz 2009 for recent reviews).  Of particular interest are
sources of coincident gravitational and electromagnetic
radiation. Besides providing unique information on the nature of the
event, such dual detections will lead to a new version of Hubble
diagram that is based on the gravitational distance measurements
(e.g. Bloom et al. 2009; Phinney 2009).

Inspirals into black holes of masses $\MBH\sim (10^5-10^6)M_\odot$
produce gravitational waves in the frequency band where LISA is most
sensitive.  Normal stars are tidally disrupted well before they
approach such black holes, and therefore discounted as possible LISA
sources (however, see Freitag 2003).  Inspirals of compact objects are
guaranteed sources of gravitational waves, however most of them are
not promising for dual detections.  In particular, stellar-mass black
hole or neutron star inspirals are not expected to generate bright
electromagnetic signals.  Only inspiraling white dwarfs (WDs) offer a
possibility for dual detection 
(Menou et. al. 2008; Sesana et al. 2008).
WDs can be tidally disrupted very
close to the black hole and then could create a transient accretion
disc with Eddington luminosity.

  Estimated EMRI rates are high enough for observations with LISA
  (e.g. Phinney 2009).  
The expected fraction of WD inpirals among all EMRIs
depends on the 
  degree
of mass segregation in galactic nuclei,
which favors stellar-mass black holes over white dwarfs in the central
cluster.  The abundance of white dwarfs also depends on the details of
stellar evolution, which are not completely understood. A fraction of
WD inpirals as large as $\sim 10\%$ has been suggested (e.g. Hopman
\& Alexander 2006a).

The orbital parameters of WD inspirals are also uncertain.  EMRIs form
when a compact object is captured onto a tight orbit whose evolution
is controlled by gravitational radiation rather than random
interactions with other stars in the central cluster around the
massive black hole. Two main channels exist for EMRI formation:
capture of single stars and capture of binary systems (e.g., Hils \&
Bender 1995; Sigurdsson \& Rees 1997; Ivanov 2002; Gair et al. 2004;
Hopman \& Alexander 2005, 2006a,b, 2007; Miller et al. 2005; Hopman
2009). In the single-capture scenario, the shrinking orbit can retain
a significant eccentricity until the end of inspiral.  In contrast,
the binary-capture scenario leads to nearly circular orbits (Miller et
al. 2005).

In this paper, we focus on WD inspirals that are not completed because
the star is tidally disrupted before its orbit becomes unstable.  We
argue that the WD begins to lose mass very gently and, for thousands
of orbital periods, this process resembles accretion through the $L_1$
point in a binary system rather than a catastrophic disruption.  This
offers a possibility of {\it simultaneous} observation of the inspiral
by LISA and traditional, optical and X-ray telescopes.  Previous work
on tidal deformation of a WD orbiting a massive black hole focused on
two extreme regimes: (i) weak deformation was studied analytically
using perturbation theory (e.g., Rathore et al. 2005; Ivanov \&
Papaloizou 2007), and (ii) strong deformation leading to immediate
disruption was simulated numerically (e.g. Kobayashi et al. 2004;
Rosswog, Ramirez-Ruiz \& Hix 2009).  The regime 
  considered in the present paper is
different from both cases explored previously.  It
involves an extended phase of strong deformation with small mass loss,
which we call `tidal stripping' below.  The mass of the WD remains
almost unchanged during this phase and 
continues to emit
gravitational waves. Even a small orbital eccentricity, e.g.
$e=0.01$, implies that tidal stripping occurs only near the pericenter
of the orbit, during a small fraction of each orbital period.  Thus,
the mass loss is expected to be {\it periodic}, possibly leading to a
periodic electromagnetic signal from the inspiral.
%
%
%
\section{Tidal stripping}
\subsection{Onset of mass loss}
Consider a WD 
orbit with semi-major axis $a$ and
eccentricity $e$. The orbital parameters gradually evolve as a result 
of gravitational radiation (Peters 1964)
\begin{eqnarray}\label{eq:adot}
 \dot a&=&-\frac{64}{5}\frac{G^{3}M \MBH^{2}}{c^{5}a^{3}(1-e^{2})^{7/2}}
     \left(1+\frac{73}{24}e^{2}+\frac{37}{96}e^{4}\right),\\ \label{eq:edot}
 \dot e&=&-\frac{304}{15}\frac{G^{3}M \MBH^{2}}{c^{5}a^{4}(1-e^{2})^{5/2}}
     \left(1+\frac{121}{304}e^{2}\right)e,
\end{eqnarray}
where dot denotes the time derivative of the secular evolution (averaged
over the orbit). Both $a$ and $e$ slowly decrease with time.
The fractional change of the pericenter radius $r_p=a(1-e)$ 
in one orbital period $P=2\pi(a^3/G\MBH)^{1/2}$ is 
\begin{equation}
\label{eq:alpha}
 \alpha \equiv\frac{-\dot r_p P}{r_p} 
    =\frac{128\pi G^{5/2}\MBH^{3/2}M}{5c^2\,r_p^{5/2}}
     \frac{\left(1-\frac{7}{12}e+\frac{7}{8}e^2+\frac{47}{192}e^3\right)}
          {(1+e)^{7/2}}.
\end{equation}
For the typical parameters of the problem considered in this paper,
$\alpha\sim 10^{-5}$. Two inaccuracies in the above formulas should be
noted: (i) Equations~(\ref{eq:adot})-(\ref{eq:alpha}) are valid only
  if $r_p\gg G\MBH/c^2$.
They become approximate during the most interesting phase of the inspiral
when $r_p$ is only a few $r_g=2G\MBH/c^2$.  
(ii) The equations neglect the effects of mass loss on the evolution of 
the orbit.\footnote{
  Let $\delta M$ be the mass lost in one orbit. The maximum correction to 
  $\alpha$ is $\sim\delta M/M$, and it is significant only when 
  $\delta M$ exceeds $\alpha M$. Section~2.3 suggests that the value of 
  $\alpha$ is unimportant at this stage, as the evolution becomes controlled 
  by the mass loss itself, not the drift of the pericenter.
}

The mass loss begins when the tidal acceleration created by the black hole
at the WD surface, $(G\MBH/r_p^3)R$, becomes comparable to $GM/R^2$, 
where $R$ is the radius of the WD. 
A similar condition for the onset of mass transfer in synchronous binary 
systems is that the donor star fills its Roche lobe.
In the more complicated case of eccentric non-synchronous binary systems, 
mass transfer may be approximately described using an instantaneous 
effective Roche lobe (Sepinsky et. al. 2007), neglecting
all the effects beyond the quasi-static limit (e.g. Ritter 1988).
We will use the following approximate condition for the mass-loss onset,
\begin{equation}
\label{eq:Rt}
   R>\Rt\equiv\gamma\; r_p\left(\frac{M}{\MBH}\right)^{1/3},
\end{equation}
where $R$ is the radius of the {\it unperturbed} star, before it
experiences any tidal deformation, and $\gamma$ is a numerical
constant.  The exact value of $\gamma$ depends on the mass ratio
$M/\MBH$ and the orbital parameters $a$ and $e$.  Even more
importantly, it also depends on the rotation of the WD and the history
of its tidal heating, none of which is known.  If $\Rt$ is interpreted
as the effective Roche-lobe studied by Sepinsky et al. (2007), their
results give $0.39 \leqslant \gamma \leqslant 0.59$, with
$\gamma=0.49$ for synchronous binary systems.  Our conclusions are
independent of the exact value for $\gamma$.  In numerical examples
where it needs to be specified, we assume  $\gamma\approx 0.5$.

\begin{figure}
\begin{center}
\includegraphics[width=2.9in]{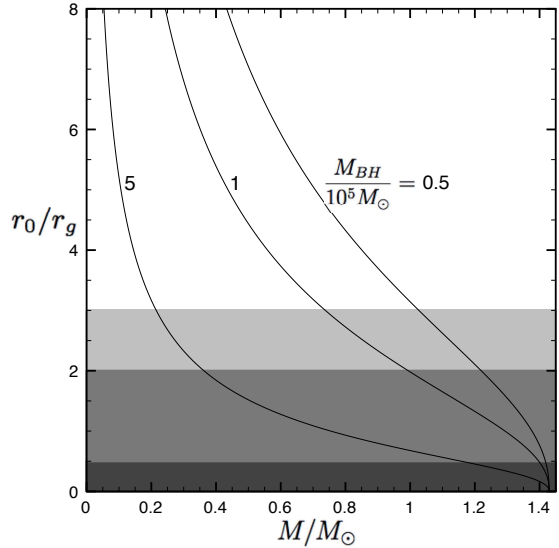}
\caption{Radius $r_0$ where tidal stripping begins, in units of the
  Schwarzschild radius
  $r_g\equiv2G\MBH/c^2\approx3\times10^{10}(\MBH/10^5M_\odot)$~cm, is
  shown as a function of the WD mass.  The three curves correspond to
  $\MBH/10^5 M_\odot=0.5$, 1 and 5.  The value of $r_0$ was estimated
  using the simplified equation~(\ref{eq:r0}), which is
  non-relativistic and neglects the effect of the black-hole spin
  $a_s$ on the tidal force. The shaded region shows radii $r<\rmin$
  where no stable bound orbits exist. 
  $\rmin$ depends on $a_s$ and orbital eccentricity $e$, in particular
$\rmin=3r_{g}$ for $\{a_s=0,e=0\}$, $\rmin=2r_{g}$ for
$\{a_s=0,e=1\}$ and $\rmin=r_{g}/2$ for $\{a_s=1,e=1\}$.  }
\label{fig_DisruptionRadius}
\end{center}
\end{figure}

The orbit slowly shrinks due to gravitational radiation and
condition~(\ref{eq:Rt}) will be first met when the pericenter radius
$r_p$ reaches the value $r_0$ estimated below.  We adopt a simple
model for the unperturbed WD star: a non-rotating cold sphere of
uniform chemical composition, supported by the pressure of degenerate
electrons. It is straightforward to obtain numerically the mass-radius
relation for such a star. We find that it is well approximated
by the following formula 
  (with less than 2 per cent error 
for $0.2M_{\odot}<M<1.4M_{\odot}$), 
\begin{eqnarray}\label{eq:R}
   R=R_\star\left(\frac{\MCh}{M}\right)^{1/3}
     \left(1-\frac{M}{\MCh}\right)^{\beta},
\end{eqnarray} 
where $ \MCh=1.43M_{\odot}$ 
is the Chandrasekhar mass, $\beta= 0.447$ and
$R_\star=0.013R_{\odot}$. Substituting $R(M)$ in
equation~(\ref{eq:Rt}), one obtains the pericenter radius at which
mass loss begins
\begin{eqnarray}\label{eq:r0}
   r_0=\frac{R_\star}{\gamma }\left(\frac{\MCh\MBH}{M^2}\right)^{1/3}
         \left(1-\frac{M}{\MCh}\right)^{\beta}.
\end{eqnarray}
Figure~\ref{fig_DisruptionRadius} shows $r_0/r_g$ 
as a function of WD mass for different $\MBH$.

%
%
\subsection{Estimate for mass loss in one orbital period}
As the star begins to lose mass, it continues to do so repeatedly: at
each pericenter passage the star overflows the Roche lobe for a short
time. The hydrodynamics of the surface layers of the tidally stripped
star is complicated. It is clear however that the mass lost in one
pericenter passage, $\dM$, will be tiny for the initial mass loss
episodes when the star surface barely touches the Roche lobe.  The
natural small parameter in the problem is $\dR/R=(R-\Rt)/R$.  The
onset of mass loss is defined by $\dR=0$ and we assume that the
dependence of $\dM$ on $\dR/R\ll 1$ can be expanded in series whose
leading term has the form,
\begin{equation}
\label{eq:dM}
  \frac{\dM}{M}=A\,\left(\frac{\dR}{R}\right)^\psi,
  \qquad \dR\equiv R-\Rt\ll R,
\end{equation} 
where $A$ is a numerical factor.

For illustration, consider a 
toy model.
Suppose that after the pericenter passage the star loses the surface shell 
$\dR\ll R$ whose mass is estimated using the structure of the unperturbed 
star,
\begin{eqnarray}\label{eq:dMtoy}
  \dM=4\pi R^2\int_0^{\dR}\rho(z)\,dz.
\end{eqnarray}
Here $z$ is the depth measured inward from the surface, and $\rho(z)$ is
the mass density of the unperturbed star at depth $z$. 
For simplicity, let us assume the polytropic
equation of state in the surface layer $P=K\rho^\gamma$ with $\gamma=5/3$,
the same as in the deeper region where electrons become degenerate.
$P$ and $\rho$ change continuously between the degenerate and non-degenerate 
regions, and therefore $K$ in the surface 
layers must be the same as for non-relativistic degenerate electron gas,
\begin{eqnarray}
   K\equiv\frac{1}{20}\left(\frac{3}{\pi}\right)^{2/3}
          \frac{h^{2}}{m_{e}(\mu_e m_p)^{5/3}},
\end{eqnarray}
where $\mu_e\approx 2$ is the mean molecular weight per electron.
The hydrostatic balance $dP/dz=\rho\,GM/R^2$ gives
\begin{eqnarray}
  \rho(z)=\left(\frac{2 G M}{5K R}\right)^{3/2}\left(\frac{z}{R}\right)^{3/2},
\end{eqnarray}
and evaluating the integral in equation~(\ref{eq:dMtoy}) one finds
$\delta M/M=A(\Delta/R)^\psi$ with
\begin{eqnarray}
\label{eq:psi}
  \psi=\frac{5}{2}, \qquad  
  A\approx 6.1\left(1-\frac{M}{\MCh}\right)^{3\beta/2}.
\end{eqnarray}
In this toy model, $\dM=(6/5)[\rho(\dR)/\bar{\rho}](\dR/R)$
where $\bar{\rho}=3M/4\pi R^3$.
The small factor $(\dR/R)^\psi\ll 1$ in $\dM$ results from the small 
thickness of the surface layer $\dR$ and the small density of this layer,
$\rho(\dR)/\bar{\rho}\sim (\dR/R)^{3/2}$.

Additional factors may enter a more realistic model. 
The duration of the mass-loss episode $\tloss$ 
may be so short that only a fraction of the $\Delta$-layer is lost.
$\tloss$ likely scales with some power of $\dR/R$. One could formulate 
a 
time-dependent model for the mass-loss episode by evaluating
$\Rt$ along the orbit around the pericenter (instead of just 
one point $r_p$). If one assumes that the mass-loss episode occurs where 
$\Rt<R$, then $\tloss\sim(\dR/R)^{1/2}\tau$ where $\tau$ is the 
sound-crossing time of the star.
A complete hydrodynamical model is three-dimensional 
as the mass loss is asymmetric: 
gas will flow through the $L_1$ point, as happens in close binary systems.
The flow velocity may be comparable to the sound speed in the outer layers 
of the star (which is smaller than the sound speed in its interior).   
Perhaps future hydrodynamical calculations will give the 
duration of the mass loss episode and the resulting $\psi$, $A$ and $\dM$.
We do not know the exact values of $\psi$ and $A$ and keep them as parameters. 
The illustrative numerical example shown 
below assumes $\psi$ and $A$ given in equation~(\ref{eq:psi}).

From equations~(\ref{eq:Rt}) and (\ref{eq:r0}) one finds
\begin{equation}
\label{eq:dR}
  \frac{\dR}{R}=1-\frac{\Rt}{R}
     =1-\frac{r_p}{r_0}\left(\frac{M}{M_0}\right)^{2/3}
         \left(\frac{\MCh-M_0}{\MCh-M}\right)^\beta.
\end{equation}
For the first mass loss episode, $M=M_0$ and $r_p/r_0\sim 1-\alpha$,
which implies $\dR /R \sim\alpha\sim 10^{-5}$.  Equation~(\ref{eq:dR})
shows that $\Delta $ grows with each passage of the pericenter as
$r_p$ decreases (due to gravitational radiation) and $M$ decreases
(due to mass loss).
\subsection{Evolution of mass loss over many orbits}
$N$ orbits after the star reached $r_0$, the pericenter radius is
given by
\begin{equation}\label{eq:rp/rp0}
  \frac{r_p}{r_0}\approx 1-\alpha N,  \qquad \alpha N \ll 1,
\end{equation}
where $\alpha$ is given by equation~(\ref{eq:alpha}). 
Here we assumed that $r_p$ decreases by $\alpha r_p$ in each orbit,
neglecting the effect of mass loss on the orbit (which is likely to 
be a poor approximation at late stages of mass loss, e.g. Bildsten \&
Cutler 1992).

Let $x$ be the mass fraction of the star that has been lost over $N$ orbits,
\begin{equation}
  x=\frac{M_0-M}{M_0}.
\end{equation}
As long as $x\ll 1$, one can expand $\Delta$ in $x$ and $\alpha N$
and keep only the leading linear terms,
\begin{equation}
\label{eq:dR1}
  \frac{\Delta}{R}\approx\frac{\Delta}{R_0} \approx \alpha\,N+B\,x, 
     \qquad B=\frac{2}{3}+\frac{\beta M_0}{\MCh-M_0}.
\end{equation}
The term $\alpha N$ describes the change in $\Rt/R_0$ due to the decreasing 
$r_p$ while the small changes in $M$ and $R$ are neglected.
The term $Bx$ describes the effect of decreasing $M$ 
on $\Rt$ and $R$ while the small drift of the pericenter is neglected. 

It is convenient to treat $N\gg 1$ as a continuous variable and describe 
the mass loss by the differential equation $dM/dN=-\dM$. 
Then substitution of equation~(\ref{eq:dR1}) to equation~(\ref{eq:dM}) 
gives the differential equation for $x(N)$,
\begin{equation}
  \frac{dx}{dN}=A\left(\alpha N+Bx\right)^\psi.
\end{equation} 
One can see from this equation that there are two stages of mass loss:
(a) $Bx\ll\alpha N$ and (b) $Bx\gg \alpha N$. Assuming $\psi>1$, we find
the solutions for the two regimes,
\begin{equation}
\label{eq:a}
  x\approx\frac{A}{\psi+1}\,\alpha^\psi N^{\psi+1}, \qquad N < N_1,
\end{equation}
\begin{equation}
\label{eq:b}
  x\approx \left[(\psi-1)AB^\psi(N_\star - N)\right]^{1/(1-\psi)}, 
        \qquad N > N_1.
\end{equation}
Equation~(\ref{eq:b}) applies only as long as $x\ll 1$, however it 
gives an estimate for the number of orbits to complete disruption 
$N_\star$.
The values of $N_1$ and $N_\star$ can be evaluated by matching $x$ and
$dx/dN$ for the two solutions at $N=N_1$,
\begin{equation}
\label{eq:N1}
  N_1=\left(\frac{\psi+1}{AB}\right)^{1/\psi}\alpha^{(1-\psi)/\psi},
\end{equation}
\begin{equation}
\label{eq:Ns}
  \frac{N_\star-N_1}{N_1}=\frac{1}{\psi^2-1}.
\end{equation}
At $N<N_1$ the growth of $x$ is caused by 
the decrease in $r_p$ at practically unchanged mass and radius of the star. 
After $N_1$ orbits, the growth of $x$ accelerates as it is now 
controlled by the decreasing $M$ (and increasing $R$) while the change 
in $r_p$ has a negligible effect. 
Equations~(\ref{eq:a}) and (\ref{eq:b}) imply that $\delta M\propto N^{\psi}$
  for $N< N_{1}$ 
and $\delta M\propto(N_{\star}-N)^{\psi/(1-\psi)}$
  for $N>N_1$.

For example, consider the toy model described by equation~(\ref{eq:psi}).
In this case, $N_1\sim \alpha^{-3/5}\sim 10^3$ and $N_\star-N_1=(4/21)N_1$.
The mass fraction lost after $N_1$ orbits is 
$x(N_1)\sim \alpha N_1/B\sim \alpha^{2/5}\sim 10^{-2}$.
The detailed behavior of $x(N)$ in the toy model is shown in Figure~2
for a WD with initial mass $M_0=0.6M_\odot$. 
The figure shows the solution of equation $dx/dN=A(\dR/R)^\psi$ with 
$\dR$ given by equation~(\ref{eq:dR}).
\begin{figure}
\begin{center}
 \includegraphics[width=2.9in]{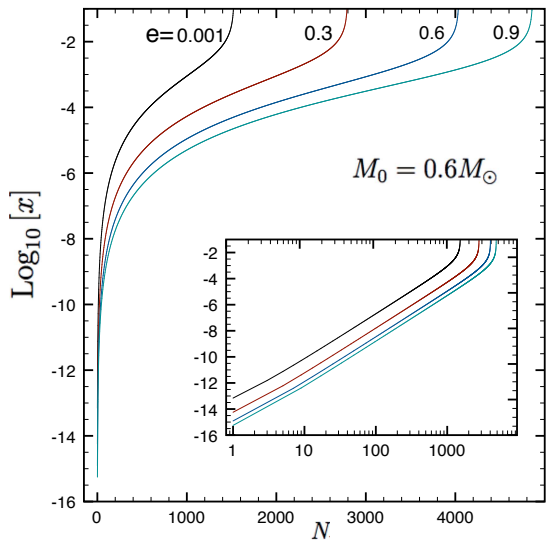}
\caption{
Lost mass fraction $x=(M_0-M)/M_0$ after $N$ orbits since the onset of tidal 
stripping. The WD has initial mass $M_0=0.6M_\odot$. 
Different curves correspond to different eccentricities of the orbit;
$\MBH=10^5M_\odot$ is assumed for all cases. 
The insert shows $\log x$ plotted against $\log N$. 
}
 \label{fig_Initial_06}
\end{center}
\end{figure}
%
%
\subsection{Periodic mass-loss rate}
The mass-loss rate $dM/dt$ is nearly periodic with the orbital period $P$.
It is zero throughout most of the orbit and has a strong peak near the 
pericenter. To illustrate this behavior, we calculated the following, 
greatly simplified model,
\begin{eqnarray} 
\label{eq:dm/dt}
  \frac{dM}{dt}=\left\{ \begin{array}{ll}
      -\dM/\tau(M)  &  \Rt<R \\
                   0        &  \Rt>R 
                        \end{array}
                 \right.
\end{eqnarray}
where $R(M)$ is the radius of the unperturbed star of mass $M$,
$\Rt$ is calculated everywhere along the orbit according to 
equation~(\ref{eq:Rt}). $\dM(t)$ is given by equation~(\ref{eq:dM})
(it is evaluated using the local value of $\Rt$);
$\tau=(G\bar\rho)^{-1/2}$ is 
  the sound-crossing time-scale of the star.

\begin{figure}
\begin{center}
\includegraphics[width=2.9in]{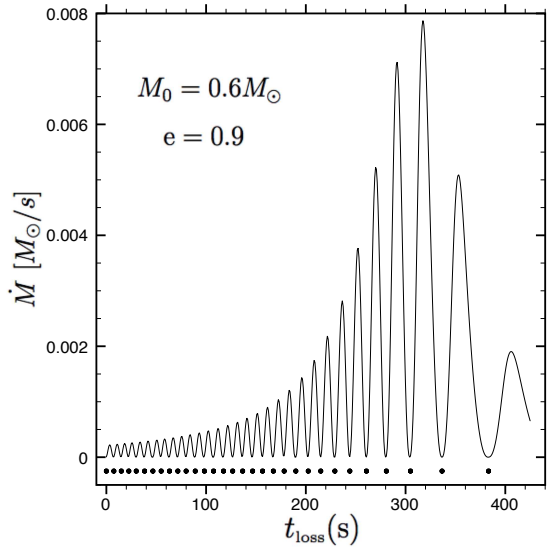}
\caption{Instantaneous mass-loss rate $\dot{M}=|dM/dt|$ 
  during the last 35 orbits before disruption. The WD with initial mass 
  $M_0=0.6M_\odot$ is orbiting
a black hole with $\MBH=10^5M_\odot$; the orbit has eccentricity $e=0.9$.
The horizontal axis shows `time during mass transfer', which 
increases only when $\dot M\neq 0$. The area under each peak is
the mass lost per pericenter passage; it follows a power law with the
number of orbits left to complete disruption, 
$\dM\propto (N_\star-N)^{-1.8}$, close to the results of Section~2.3.
}
\label{fig_Mdot}
\end{center}
\end{figure}

Figure~\ref{fig_Mdot} shows the numerical solution of
equation~(\ref{eq:dm/dt}) for the last 
  35
orbits before
disruption. The WD is assumed to have an initial mass
$M_0=0.6M_{\odot}$ and orbital eccentricity $e=0.9$;
the orbital period is $P\approx 2\times 10^3$~s.  We plot
the instantaneous mass-loss rate 
  $|dM/dt|$ 
versus time $t_{\rm loss}$
shown by a clock that ticks only when $dM/dt\neq 0$.  The periodic peaks in 
  $|dM/dt|$ 
coincide with the pericenter passages.
%
%
%
%
\section{Discussion: electromagnetic counterpart}
The standard picture of EMRI envisions an orbit that gradually shrinks 
due to gravitational radiation until its pericenter $r_p$ reaches $\rmin$ 
where the orbit becomes unstable and plunges into the black hole.
The main observational phase of the inspiral 
is when $r_p$ decreases from $\sim 2\rmin$ to $\rmin$.
 For WDs with mass $M\sim M_\odot$ inspiraling into a black hole with 
 $\MBH\sim 10^5M_\odot$ 
the main observational
phase lasts $\sim 10^5$ orbital periods, which may be comparable to one 
year, depending on the orbital eccentricity. 
We argued in Section~2 that the inspiraling WDs can experience an 
extended period of slow mass loss during observations by LISA.

To summarize, the tidal stripping begins very gently because the
pericenter of the orbit drifts inward slowly, by a tiny fraction
$\alpha\sim 10^{-5}$ in one orbital period. This leads to many
repeated episodes of small mass loss. The process 
  occurs
at radius $r_0$ (eq.~\ref{eq:r0}), comparable to $\rmin$.  The star is 
strongly deformed by the tidal forces near $r_0$ but it barely touches 
its Roche lobe for a small fraction of the orbital period. As a result,
the star loses a small amount of mass through the $L_1$ point at each
pericenter passage. To our knowledge, this regime of tidal stripping
was 
  not explored by
direct hydrodynamical simulations.  Our crude
estimates suggest two phases of stripping: the first phase lasts
$N_1\sim 10^3$ orbits until the star loses $\sim 1$ per cent of its
mass.  Then the mass loss accelerates: the decrease in $M$ (and the
corresponding increase in the WD radius) implies that the star
overfills the Roche lobe more and more with every pericenter
passage. As a result, the star loses the remaining 99 per cent of its
mass in a few hundreds of additional orbits.  
The mass lost in each
individual episode during these last hundreds of orbits behaves as
$\dM\propto (N_\star-N)^{-\xi}$ where $N_\star-N\ll N_\star$ is the number
of orbits remaining to complete disruption and $\xi\sim \psi/(\psi-1)$.

According to our estimates, tidal stripping operates for 
  days or weeks,
creating a relatively long-lived source of gas.  The 
gas can
accrete onto the black hole and produce significant electromagnetic
radiation together with the gravitational waves observed by LISA.  The
radiation source may become bright before the mass loss spoils the
standard EMRI template for the gravitational-wave signal.

The gas produced by tidal stripping moves on nearly Keplerian orbits
that are initially close to the WD orbit. 
  The gas probably leaves the star with relative velocity 
  $\sim v_{\rm esc}=(2GM/R)^{1/2}$ and its orbital energy differs 
  from that of the star by a small fraction 
  $\sim v_{\rm esc}/v\sim(Mr_0/\MBH R)^{1/2}
   \sim\gamma^{-1/2}(M/\MBH)^{1/3}\sim 1/30$. 
  After $\sim 30$ orbital periods, the
differential rotation of the gas has stretched it into a ring around
the black hole.  A non-zero eccentricity of the donor orbit will
create an eccentric ring.  It should viscously spread and accrete onto
the black hole.  A small mass-loss fraction $x$ can be a huge source
of gas for accretion.  If most of the stripped matter is accreted by
the black hole, the accretion rate is $\dot{M}\sim \dM/P$. It exceeds
the Eddington value $\dot{M}_{\rm Edd}\sim
10^{23}(\MBH/10^5M_\odot)$~g~s$^{-1}$ after $N\sim 10$ orbits since
the beginning of tidal stripping, well before the final disruption of
the WD. The accretion timescale in the viscous ring can be estimated
as $\tacc\sim \alpha_v^{-1}(H/r)^{-2}P$ where $P$ is the WD orbital
period, $\alpha_v=0.01-0.1$ is the viscosity parameter and $H$ is the
thickness of the ring (e.g. Shakura \& Sunyaev 1973).  $H/r\sim 1$ is
expected for super-Eddington accretion, which leads to $\tacc\sim
(10-100)P$.

This suggests that a bright source with Eddington luminosity 
$L_{\rm Edd}\sim 10^{43}$~erg~s$^{-1}$ is created 
  quickly, in less than 1 day
after the
beginning of tidal stripping. For a typical distance to such LISA
sources, $d\sim 100$~Mpc, the accretion ring should be detectable with
optical and X-ray telescopes provided its approximate location on the
sky is known.  LISA is expected to localize EMRIs within $\sim
10$~deg$^2$ (Barack \& Cutler 2004). For massive black-hole mergers,
the localization information will be available weeks to months prior
to the final coalescence (Kocsis et al. 2007, 2008; Lang \& Hughes
2008), and the localization expectations for WD EMRIs are similar
(S. Drasco, J. Gair, I. Mandel, E. Porter, private communications),
giving sufficient time for simultaneous optical and X-ray observations
during inspiral.

Mass transfer in WD inspirals is special as it creates a long-lived
source of gas very close to the black hole horizon, $r_0-r_g\sim r_g$.
As a result, the donor orbit is generally not confined to a plane, if
the black hole rotation is significant.  Besides, the orbit will
experience fast precession.  Therefore, the freshly stripped gas may
collide with the previously released gas and generate shocks
(e.g. Evans \& Kochanek 1989).  The resulting pattern of accretion may
be complicated and needs 
careful study.

An 
  intriguing
feature of tidal stripping is the periodic supply of gas.
It may leave a fingerprint on the observed luminosity, modulating 
it with the WD orbital period $P$.
The modulation of $\dot{M}$ may create a detectable 
oscillation in the luminosity from the accreting ring, even though the 
accretion timescale 
  $\tacc\sim (10-100)P\gg P$,
 tends to reduce the amplitude of modulation.
The shock emission from 
collisions between the periodic flow from the $L_1$ point and the gas 
accumulated around the black hole may be strongly modulated.

The description of mass transfer in this paper is greatly simplified.
The possibility of many repeated mass-transfer episodes is robust, but
the exact rate of tidal stripping and the dynamics of accretion need
to be explored with dedicated numerical simulations.  The great
potential that such events hold for joint detections of gravitational
and electromagnetic radiation provides motivation for 
  the effort.
The joint detection would let us witness, in real-time and
unprecedented detail, the slow tidal stripping of a WD followed by its
complete disruption.
 Note that LISA observations are expected to provide the mass and spin of 
 the black hole, as well as the details of the inspiral orbit. This can be 
 used to model in detail the hydrodynamics of accretion.

The scenario discussed in this paper assumes $r_0>\rmin$ and uses a
semi-Newtonian description for the WD orbit. A fully relativistic
model will be needed to accurately evaluate the parameter space for
such events.  The relativistic effects are especially important if the
black hole is rapidly rotating -- then $r_0$ and $\rmin$ will depend
on the black hole mass $\MBH$, its spin parameter $a_s$, the
eccentricity of the orbit, and the angle between the orbital angular
momentum and the angular momentum of the black hole. The competition
between tidal disruption and gravitational capture by rotating black
holes was investigated for parabolic orbits in Beloborodov et
al. (1992). 
  In a broad range of $\MBH$, 
the fate of a star approaching the black hole depends on the orbit 
orientation and can be either disruption or capture. For inspirals with 
$r_g<r_0<\rmin$, tidal stripping does not occur. Instead, $r_p$ reaches 
$\rmin$ and the orbit loses stability before any mass is lost by the WD. 
Then the star is crushed by tidal forces as it falls into the black hole. 
This strong and immediate disruption is different from the gentle stripping
considered here, suggesting rich phenomenology of WD inspirals.
\section*{Acknowledgements}
This work was supported in part by NASA ATFP grant NNXO8AH35G and by
the National Science Foundation under Grant No. PHY05-51164.


\begin{thebibliography}{3}

\bibitem[\protect\citeauthoryear{Bildsten}{1992}]{B92}
Bildsten L., Cutler C., 1992, ApJ, 400, 175	

\bibitem[\protect\citeauthoryear{Barack}{2004}]{BC04}
Barack L., Cutler C., 2004, Phys. Rev. D, 69, 082005

\bibitem[\protect\citeauthoryear{B}{1992}]{B92}
Beloborodov A.~M., Illarionov A.~F., Ivanov P.~B., Polnarev A.~G.,
1992, MNRAS, 259, 209

\bibitem[\protect\citeauthoryear{Bloom}{2009}]{BC04} Bloom J. S., Holz
  D. E., Hughes S. A., Menou K., 2009, Astro2010 Science White Papers,
  no. 20 (arXiv:0902.1527)

\bibitem[\protect\citeauthoryear{Cutler}{1998}]{C98}
Cutler C., 1998, Phys. Rev. D, 57, 7080

\bibitem[\protect\citeauthoryear{Evans et. al.}{1989}]{E89}
Evans C. R., Kochanek C. S., 1989, ApJ, 346, L13

\bibitem[\protect\citeauthoryear{Freitag}{2003}]{F03}
Freitag M., 2003, ApJ, 583, L21

\bibitem[\protect\citeauthoryear{Gair et. al.}{2004}]{G04}
Gair J. R. et al., 2004, Class. Quant. Grav., 20,  S1595

\bibitem[\protect\citeauthoryear{Hils et al.}{1995}]{H95}
Hils D., Bender P. L., 1995, ApJ, 445, L7	

\bibitem[\protect\citeauthoryear{Hopman}{2009}]{H09}
Hopman C., 2009, Class. Quant. Grav., 26, 094028

\bibitem[\protect\citeauthoryear{Hopman et al.}{2005}]{H05}
Hopman C., Alexander T., 2005, ApJ, 629, 362

\bibitem[\protect\citeauthoryear{Hopman et al.}{2006a}]{H06a}
Hopman C., Alexander T., 2006a, ApJ, 645, L133

\bibitem[\protect\citeauthoryear{Hopman et al.}{2006b}]{H06b}
Hopman C., Alexander T., 2006b, ApJ, 645, 1152

\bibitem[\protect\citeauthoryear{Hughes}{2009}]{H09}
Hughes S. A., 2009, ARA\& A, 47, 107

\bibitem[\protect\citeauthoryear{Ivanov}{2002}]{I02}
Ivanov P. B., 2002, MNRAS, 336, 373

\bibitem[\protect\citeauthoryear{Ivanov et al.}{2007}]{I07}
Ivanov P. B., Papaloizou J. C. B., 2007, A\& A, 476, 121

\bibitem[\protect\citeauthoryear{Kobayashi et al.}{2004}]{K04}
Kobayashi S., Laguna P., Phinney E. S., Meszaros P., 2004, ApJ, 615, 855

\bibitem[\protect\citeauthoryear{Kocsis et al.}{2007}]{K07}
Kocsis B., Haiman Z., Menou K., Frei Z., 2007, Phys. Rev. D, 76, 022003

\bibitem[\protect\citeauthoryear{Kocsis et al.}{2008}]{K08}
Kocsis B., Haiman Z., Menou K., 2008, ApJ, 684, 870

\bibitem[\protect\citeauthoryear{Lang et al.}{2008}]{L08}
Lang R. N., Hughes S. A., 2008, ApJ, 677, 1184

\bibitem[\protect\citeauthoryear{Menou et al.}{2008}]{MHK08}
Menou K., Haiman Z., Kocsis B., 2008, NewAR, 51, 884

\bibitem[\protect\citeauthoryear{Miller et al.}{2005}]{Miller05}
  Miller, M. C., Freitag, M., Hamilton, D. P. \& Lauburg, V. M. 2005,
  ApJ, 631, L117

\bibitem[\protect\citeauthoryear{Peters}{1964}]{P64}
Peters P. C.,  1964, Phys. Rev., 136, B 1224

\bibitem[\protect\citeauthoryear{Phinney}{2009}]{P09} Phinney E. S.,
  2009, Astro2010 Science White Papers, no. 235 (arXiv:0903.0098)

\bibitem[\protect\citeauthoryear{Rathore et. al.}{2005}]{R05}
Rathore Y., Blandford R. D., Broderick A. E., 2005, MNRAS, 357, 834

\bibitem[\protect\citeauthoryear{Ritter}{1988}]{R88}
Ritter H., 1988, A\&A, 202, 93

\bibitem[\protect\citeauthoryear{Rosswog et. al.}{2009}]{R09}
Rosswog S., Ramirez-Ruiz E., Hix W. R., 2009, ApJ, 695, 404

\bibitem[\protect\citeauthoryear{Sathyaprakash}{2009}]{S09}
Sathyaprakash B. S., Schutz B. F., 2009, Liv. Rev. Rel., 12, 2

\bibitem[\protect\citeauthoryear{Sepinsky}{2007}]{S07}
Sepinsky J. F., Willems B., Kalogera V., 2007, ApJ, 660, 1624

\bibitem[\protect\citeauthoryear{Sesana}{2008}]{S08}
Sesana A., Vecchio  A., Eracleous M., Sigurdsson S., 2008, MNRAS, 391, 718

\bibitem[\protect\citeauthoryear{Shakura}{1973}]{SS73}
Shakura N.~I., Sunyaev R.~A., 1973, A\&A, 24, 337

\bibitem[\protect\citeauthoryear{Sigurdsson et. al.}{1997}]{S97}
Sigurdsson S., Rees M. J., 1997, MNRAS, 284, 318

\end{thebibliography}
\end{document}